\let\latexaddtocontents\addtocontents
\documentclass[submitting, floatfix]{nst}
\let\addtocontents\latexaddtocontents
\usepackage{natbib}
\usepackage{subfigure,dcolumn}
\usepackage[T2A,T1]{fontenc}
\usepackage[english]{babel}
\usepackage{color}
\usepackage{listings}
\usepackage{textcomp,mathcomp}
\usepackage{comment}
\usepackage{lineno}

\let\footnote\relax

\let\textcite\relax

\let\citeauthor\relax
\let\citeyear\relax
\expandafter\let\csname
ver@natbib.sty\endcsname\relax
\let\linenumbers\relax

\usepackage[
    backend=biber,
    bibencoding=utf8,
    style=numeric-comp,
    giveninits=true,
    autolang=other, 
    sorting=none,
    doi=true, isbn=false, eprint=false, url=false, date=year, 
    maxcitenames=3,
    mincitenames=3,
    maxbibnames=3,
    minbibnames=3,
]{biblatex}
\usepackage{csquotes}
\renewbibmacro{in:}{}
\begin{document}
\renewcommand{\bibliography}[1]{}

\title{A ROOT based detector geometry and event visualization system for JUNO-TAO}

\author{Ming-Hua Liao}
%\affiliation{School of Physics, Sun Yat-Sen University, Guangzhou 510275, China}
\author{Kai-Xuan Huang}
\affiliation{School of Physics, Sun Yat-Sen University, Guangzhou 510275, China}
\author{Yu-Mei Zhang}
\email[Corresponding author, ]{zhangym26@mail.sysu.edu.cn}
\affiliation{Sino-French Institute of Nuclear Engineering and Technology, Sun Yat-Sen University, Zhuhai 519082, China}
\author{Jia-Yang Xu}
\author{Guo-Fu Cao}
\affiliation{Institute of High Energy Physics, Chinese Academy of Sciences, Beijing 100049, China}
\author{Zheng-Yun You}
\email[Corresponding author, ]{youzhy5@mail.sysu.edu.cn}
\affiliation{School of Physics, Sun Yat-Sen University, Guangzhou 510275, China}

\begin{abstract}
The Taishan Antineutrino Observatory~(TAO or JUNO-TAO) is a satellite experiment of Jiangmen Underground Neutrino Observatory~(JUNO) and located near the Taishan nuclear power plant~(NPP). TAO will measure the energy spectrum of reactor antineutrinos with unprecedented precision, which will benefit both reactor neutrino physics and the nuclear database.
A detector geometry and event visualization system has been developed for TAO. 
The software is based on ROOT packages and embedded in the TAO offline software framework. 
It provides an intuitive tool to visualize the detector geometry, tune the reconstruction algorithm, %scan the rare events in data analysis, 
understand the neutrino physics, and monitor the operation of reactors at NPP. 
The further applications of the visualization system in the experimental operation of TAO and its future development are also discussed.
%The software is developed dependently on the JUNO offline software and doesn't conflict with the JUNO event display software. Users can efficiently run the event display software SERENA in the JUNO offline software.

\end{abstract}


\maketitle
\linenumbers
\section{Introduction}
\label{sec:intro}

% Introducing TAO experiment and EventDisplay based on ROOT
The Taishan Antineutrino Observatory~(TAO, also known as JUNO-TAO)~\cite{JUNO:2020ijm}, a satellite experiment of Jiangmen Underground Neutrino Observatory~(JUNO)~\cite{JUNO:2021vlw}, is currently under construction in Guangdong, China. TAO is a ton-level liquid scintillator~(LS) detector and will detect anti-neutrino via the inverse beta decay~(IBD) process. Its central detector will operate at -50~$\tccentigrade$ to reduce the dark noise of Silicon Photomultiplier~(SiPM) to an acceptable level~\cite{Sanfilippo:2022gvf, Xie:2020bqa, Anfimov:2020ikk, Wang:2020typ}. Located near the Taishan nuclear power plant~(NPP), JUNO-TAO will provide a model-independent reference spectrum for reactor antineutrino measurement to JUNO with an energy resolution about 2$\%$ at 1 MeV~\cite{JUNO:2015zny, Lombardo:2023rqj, Lombardo:2024vbh, JUNO:2022mxj, Capozzi:2020cxm, JUNO:2024jaw}, which approaches the energy resolution limit of LS detector. Furthermore, JUNO-TAO aims to provide benchmark measurements for nuclear databases~\cite{JUNO:2021vlw}, search for sterile neutrinos~\cite{Steiger:2022fhu, Berryman:2021xsi}, and verify the technology for reactor monitoring to enhance the reliability of NPP operation and safeguard~\cite{LOMBARDO2023168030, Bowden:2008gu, NUCIFER:2015hdd, DayaBay:2017jkb}.

Event display is a visualization tool that illustrates detector geometry and event data in all phases of any high-energy physics~(HEP) experiment~\cite{Bellis:2018hej, HEPSoftwareFoundation:2017ggl}. It is used to design detectors, check the detector structure and elements, help with problem diagnosis with online monitoring during data taking, and understand physics in an auxiliary way in data analysis. The event display tool also allows physicists to understand simulation and reconstruction performance when tuning the reconstruction algorithm in software development~\cite{Li:2021oos}.

A ROOT-based event display has been developed for JUNO-TAO. It includes the ROOT package Event Visualization Environment~(EVE), which provides an intuitive way to construct event display~\cite{Brun:1997pa, Tadel:2008zz, Tadel:2010zz, Lin:2017usg}. The EVE was first developed in the ALICE offline project~\cite{ALICE:2008ngc} and has been widely applied in many HEP experiments~\cite{Kovalskyi:2010zz, CMS:2008xjf}.
%For example, the Fireworks~\cite{Kovalskyi:2010zz}, which is developed with the EVE package, has been used for the event display of CMS detector~\cite{CMS:2008xjf}. 
In addition to the primary visual display functions of the detector geometry and event, the software also supports the functions of 2D hits projection, data association from different formats, and graphical user interface~(GUI) control for users to check the detector geometry and event data interactively. Thus, it is an effective tool for the TAO experiment to understand neutrino physics and monitor the reactors' daily operation at NPP. 

The rest of this paper is structured as follows: Section~\ref{sec:scheme} briefly describes the software structure and data flow of the TAO event display. Section~\ref{sec:vis} presents the visualization functions being realized. Section~\ref{sec:fea}~and~Section~\ref{Perfor} show the features and performance of the TAO event display and its applications. Finally, Section~\ref{sec:con} gives the summary.
%and introduces the future development. 

% updates in this work
%%------------------------------------------
\section{Methodologies}
\label{sec:scheme}

In this section, we illustrate the structure and data flow in the visualization software of TAO.
The detector geometry and its description, as well as the event data model in offline software, will be introduced.
The construction and functional overview of the GUI interface to realize user interaction will also be expounded.

\subsection{Software structure and data flow}
\label{sec:structure}

An event display software should provide visualization functions of the detector geometry, display different views of detectors, distribution of event hits, and interactive control between the display and the users. %the differences between simulation and reconstruction from the same events information, 
Some similar geometric shapes can be selected to replace the complex detector components. 
For the JUNO experiment, the Software for Non-collider Physics Experiments~(SNiPER)~\cite{Zou:2015ioy}, which is a light-weight flexible framework relying on external packages Geant4~\cite{GEANT4:2002zbu} and ROOT~\cite{Brun:1997pa}, is used to construct the offline software system. 
JUNO is based on the SNiPER framework to develop the visualized tools~\cite{You:2017zfr, Zhu:2018mzu}. 
The same framework has been constructed for TAO to meet the requirement of consistent software architecture. 
%The architecture and data flow of the TAO event display software are shown in Fig.~\ref{fig:scheme}.

%--fig: the structure of software
\begin{figure*}[htbp]
    \centering
    \includegraphics[width=0.70\textwidth]{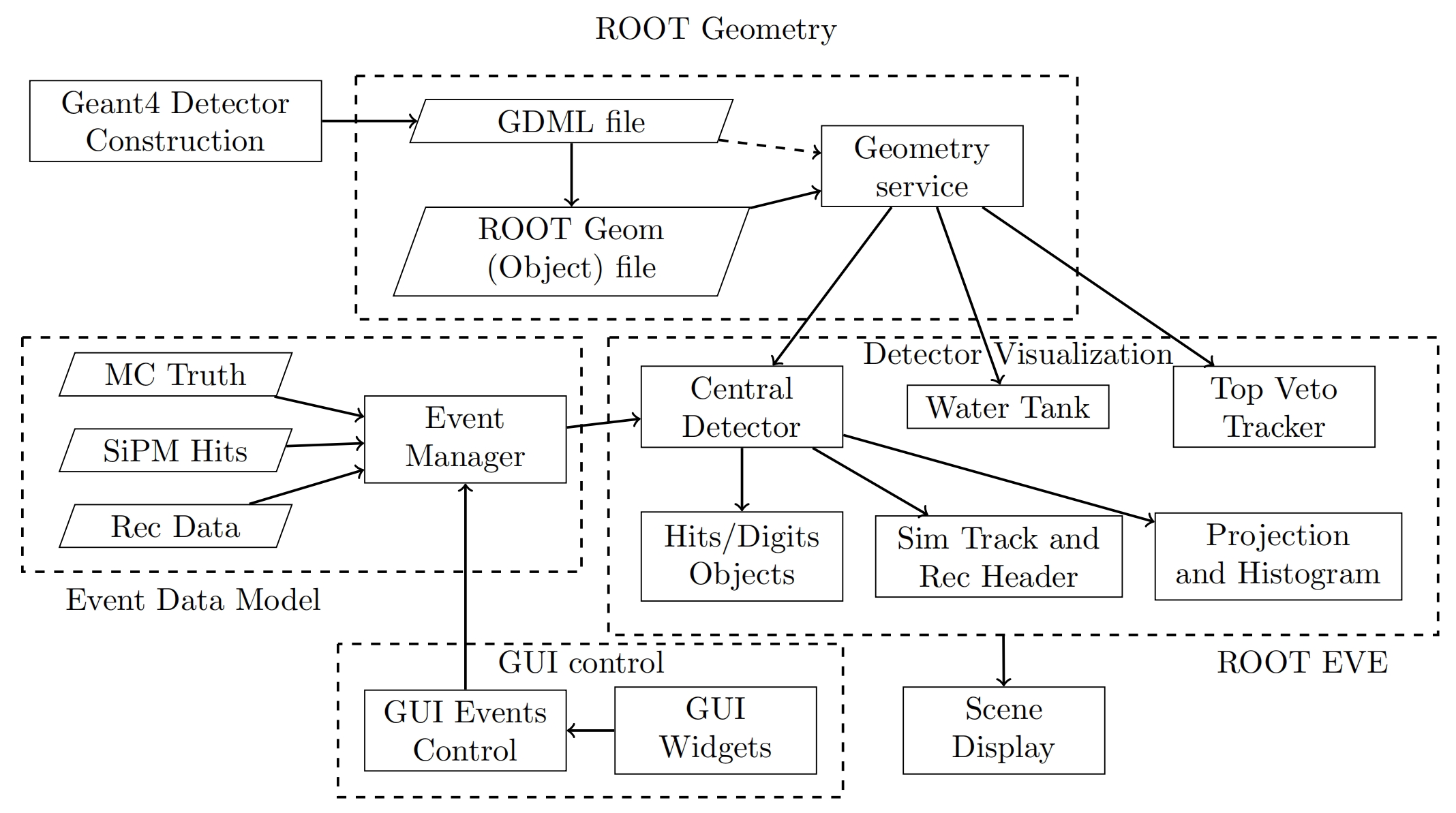}{\centering}
    \caption{The architecture and data flow of the TAO event display software. %The structure of event display software of JUNO-TAO. The geometric information of the detector simulated by Geant4 is stored in a GDML file and transformed to store the geometry information in a ROOT file. The geometric information of the detector components in the ROOT file is accessed through the Geometry service. The Event Data Model includes simulation, calibration, and reconstruction information, which can be read through the Event Manager. The ROOT EVE package in the middle is implemented to gather information that needs to be visualized and form the scene being displayed. The ROOT GUI package is used to control the events and display interface.
    }
    \label{fig:scheme}
\end{figure*}

As shown in Fig.~\ref{fig:scheme}, the TAO event display software is composed of four parts. 
\begin{itemize}
\item \emph{Detector geometry}. 
The geometric details of the detector, constructed with Geant4 in detector simulation, are initially saved in a Geometry Description Markup Language~(GDML)~file~\cite{GDML, Li:2018fny, Zhang:2020jkg}. The GDML file is then transformed into a ROOT file with the GDML-ROOT converter. The geometric information of the detector components within the ROOT file is then accessed via the geometry service~\cite{You:2008, Liang:2009zzb} and used for the 3D and 2D construction of detector units with the ROOT EVE package. %The visualization of sub-detectors includes the Silicon Photomultiplier~(SiPM) of the Central detector (CD), Water Tank (WT) and Photomultiplier Tubes (PMT), Top Veto Tracker (TVT) which is composed of the four-layer plastic scintillator (PS). The specific details about the detector display will be introduced in the next section. 

\item \emph{Event data model}. 
The Event Data Model~(EDM)~\cite{Li:2017zku} contains three types of information: simulation, calibration, and reconstruction, which is generated by TAO offline software. The EDM information is then read by the event manager and displayed together with detector geometry. The detailed information on the hits, tracks, and their associations are attributed to the corresponding objects and viewed with the information according to the requirements of users.
%That only needs to click the corresponding SiPM tile with the cursor. The SiPM of hits is projected into the 2D plot to help users better observe the distribution of the hits in the detector. 
\item \emph{EVE based visualization}. 
%Third, Event Visualization Environment is a package in ROOT. 
The detector geometry objects, the information of simulation and reconstruction, hit objects, and histograms are implemented with the corresponding visualization objects in the ROOT EVE package. %It also has the function of controlling the color of the visualized page, space size, and so on. 
The information of the detector units, including its identifier, position, time and charge of a hit, is also attached to the EVE visualization objects for interactive display.
\item \emph{GUI}. 
With the interactive functions realized by the ROOT GUI widgets, the users have access to control of the visualization objects, such as event navigation, switching between different views of the detector, and animation of the event propagation with time. These functions help the users better understand the physics processes in data processing and analysis. 
\end{itemize}

All these functions are realized with the ROOT packages and within the SNiPER framework, which makes them easy to implement in the TAO offline software and can be naturally accessed at different stages of data processing, including simulation, reconstruction, and data analysis.

\subsection{The TAO detector and geometry}

%--fig: the detector of TAO from Conceptual Design Report
\begin{figure*}[htbp]
	\centering
	\includegraphics[width=0.8\textwidth]{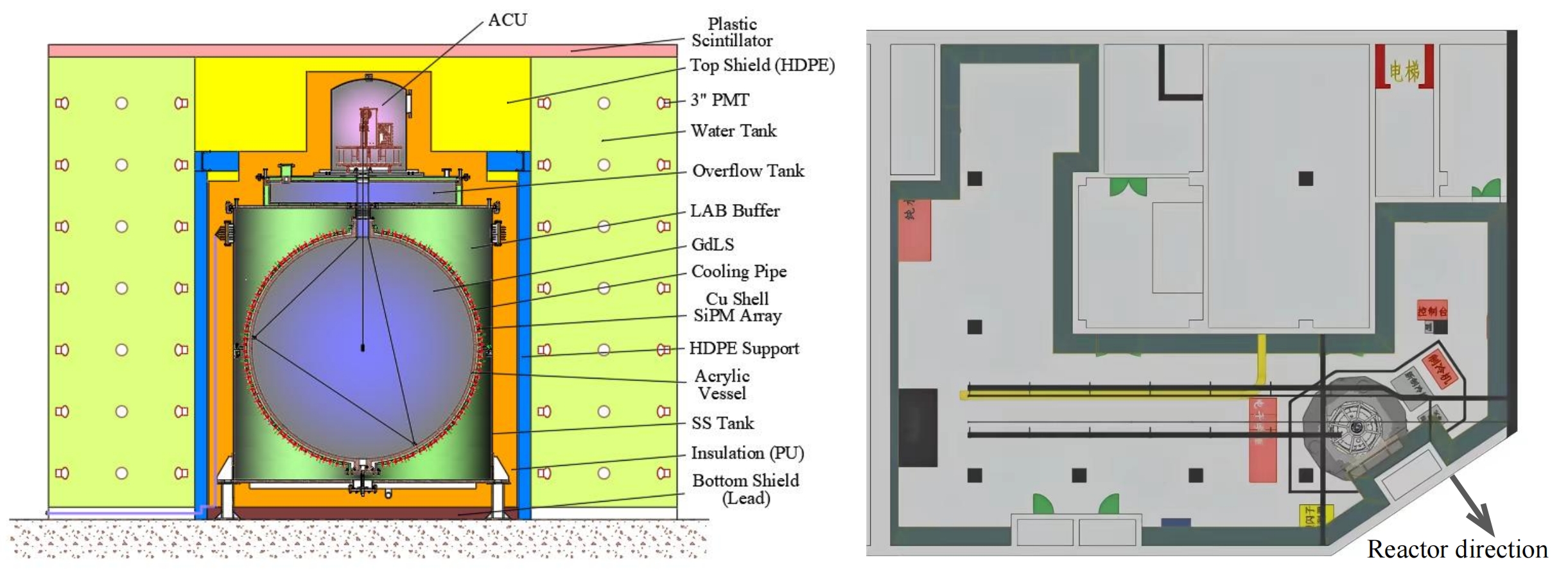}{\centering}
    \caption{Schematic view of the TAO detector. The left figure is the designed structure of the TAO detector. The right figure shows the Taishan Neutrino Laboratory. The footprint of the TAO detector is the black prismatic block located in the bottom right, which is about 44 meters from the reactor core~\cite{JUNO:2020ijm}. %The footprint of the TAO detector is the pink blocks which are about 30 m from the reactor core in the north-west direction. %Reproduced from ~\cite{JUNO:2020ijm} with permission.
    }
	\label{fig:TAOdetector_design_report}
\end{figure*}

%The TAO detector is located about 30 meters away from one of the reactor cores at the Taishan NPP~(4.6~GW). 

The TAO detector is located at a distance of approximately 44 meters from one reactor core~(TS-C1) of the Taishan NPP and 217 meters from the other reactor core~(TS-C2)~\cite{JUNO:2024jaw}. It consists of three main sub-detectors: Central Detector~(CD), Water Tank~(WT), and Top Veto Tracker~(TVT), as shown in the schematic structure of the TAO detector in Fig.~\ref{fig:TAOdetector_design_report}.

The CD is a spherical acrylic vessel filled with 2.8 tons of gadolinium-doped LS. 
It is supported with a spherical copper shell installed with about 4,100 SiPM tiles~\cite{Yu:2022god}, which has a geometric coverage of 94\% and exhibits outstanding energy resolution. 
The residual non-uniformity, degradation in energy resolution, and bias are maintained at the level of 0.2$\%$, 0.05$\%$, and 0.3$\%$ through calibration, respectively~\cite{Xu:2022mdi}. 
Thus, it can provide a reference spectrum for measuring the neutrino mass ordering in JUNO and eliminate potential model-dependent systematic uncertainties from the fine structure in the reactor neutrino spectrum~\cite{JUNO:2015zny}. 

To veto the cosmic ray muons and the radioactive backgrounds from the environment, the CD is surrounded by the WT, whose interior wall is instrumented with arrays of about 300 3-inch Photomultiplier Tubes~(PMT) for recording the information of cosmic muons via the detection of Cherenkov light. 
The TVT, which is composed of four layers of plastic scintillator, is designed to detect the tracks and direction of cosmic muons from the top of CD and WT~\cite{Luo:2023inu}, which is necessary for the ground-based detector.
%The water tank and top veto tracker form the veto detector to shield the central detector. 
More details about the TAO detector are available in Ref.~\cite{JUNO:2020ijm}. 

%The visualization of the TAO detector is applied to extracting 
The detector geometry in TAO offline software is constructed at the stage of Geant4-based detector simulation. The necessary detector information from the simulation is then exported in GDML format, including the CD, WT, and TVT. The GDML file is then converted to ROOT format and used as input for the detector geometry service, providing the storage of detector unit information conveniently and consistently, such as the position and orientation of each SiPM, the position of each plastic scintillator, the transformation between the local coordinate of every PMT and the global coordinate of TAO.

%Fig.~\ref{fig:gdml_tao} shows the geometric structure of the TAO detector simulated by Geant4, which is stored in the ROOT file and displayed with the ROOT OpenGL viewer. 

\begin{figure}[htbp]
	\centering
	\includegraphics[width=0.4\textwidth]{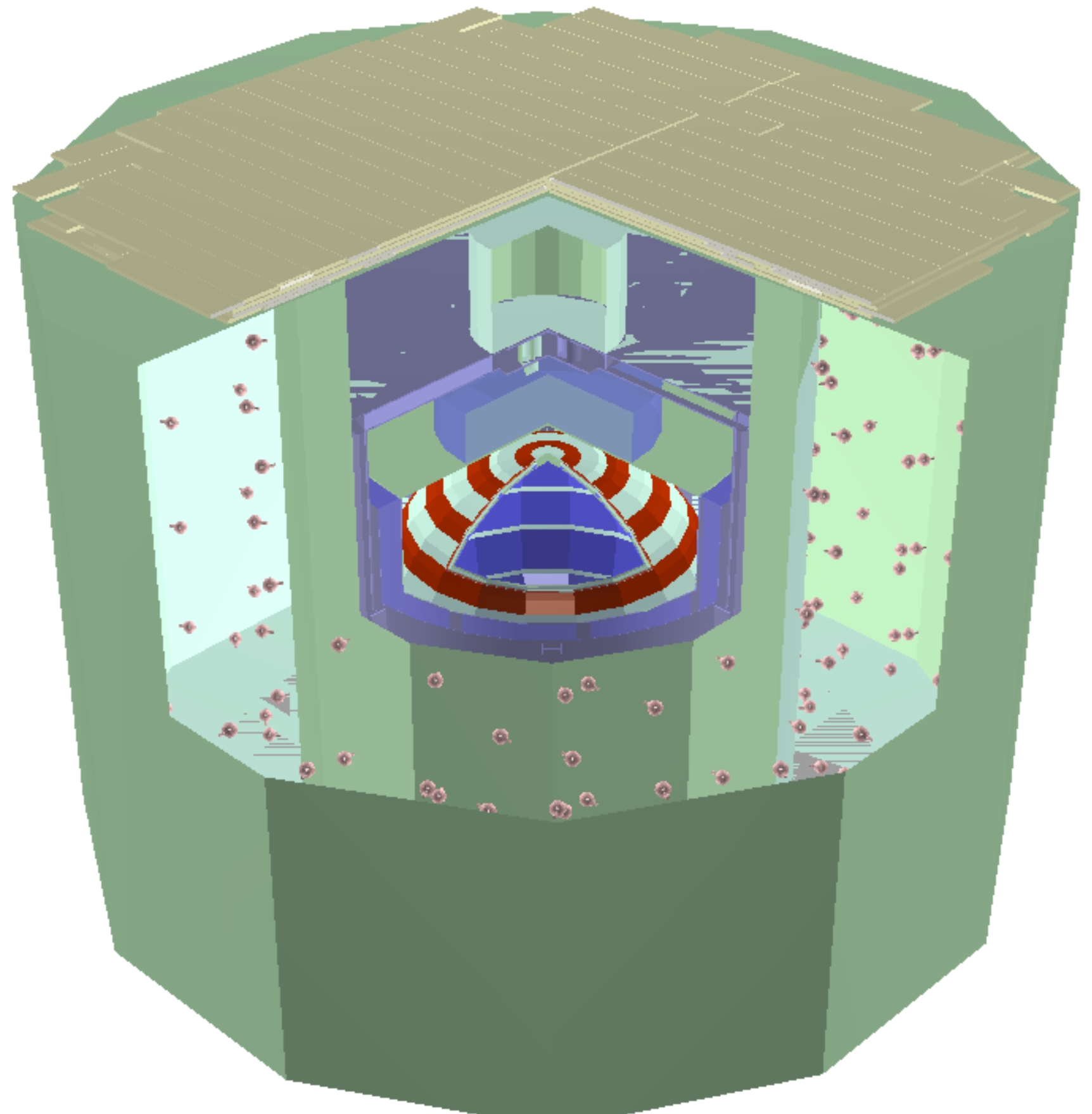}{\centering}
    \caption{The geometric structure of the TAO detector exported from GEANT4 simulation and displayed with ROOT OpenGL viewer. The brown top layers are TVT, the sphere in the center is CD, and the outer perimeter consists of PMTs installed in WT.}
	\label{fig:gdml_tao}
\end{figure}

The geometric structure of the TAO detector used in simulation and reconstruction is shown in Fig.~\ref{fig:gdml_tao}, which is stored in the GDML file format and displayed with ROOT OpenGL~\cite{Tadel:2008zzb} viewer. 

In the TAO experiment, the GDML files are converted to ROOT format, which makes it easy for users to learn about the different sub-detector structures with the ROOT browser while retaining the advantages of GDML files. The component information of JUNO-TAO is also stored in the ROOT file. All displayed objects are controlled by ROOT EVE geometry objects and are initialized with visualization attributes for later display control. With the detector data exported by the geometry service, visualization of the sub-detectors can be quickly constructed with some basic 3D models of geometric shapes, such as boxes, cones, spheres, and cylinders.

\subsection{Event data model}
%--fig: the detector of TAO from Conceptual Design Report
\begin{figure*}[htbp]
	\centering
	\includegraphics[width=0.85\textwidth]{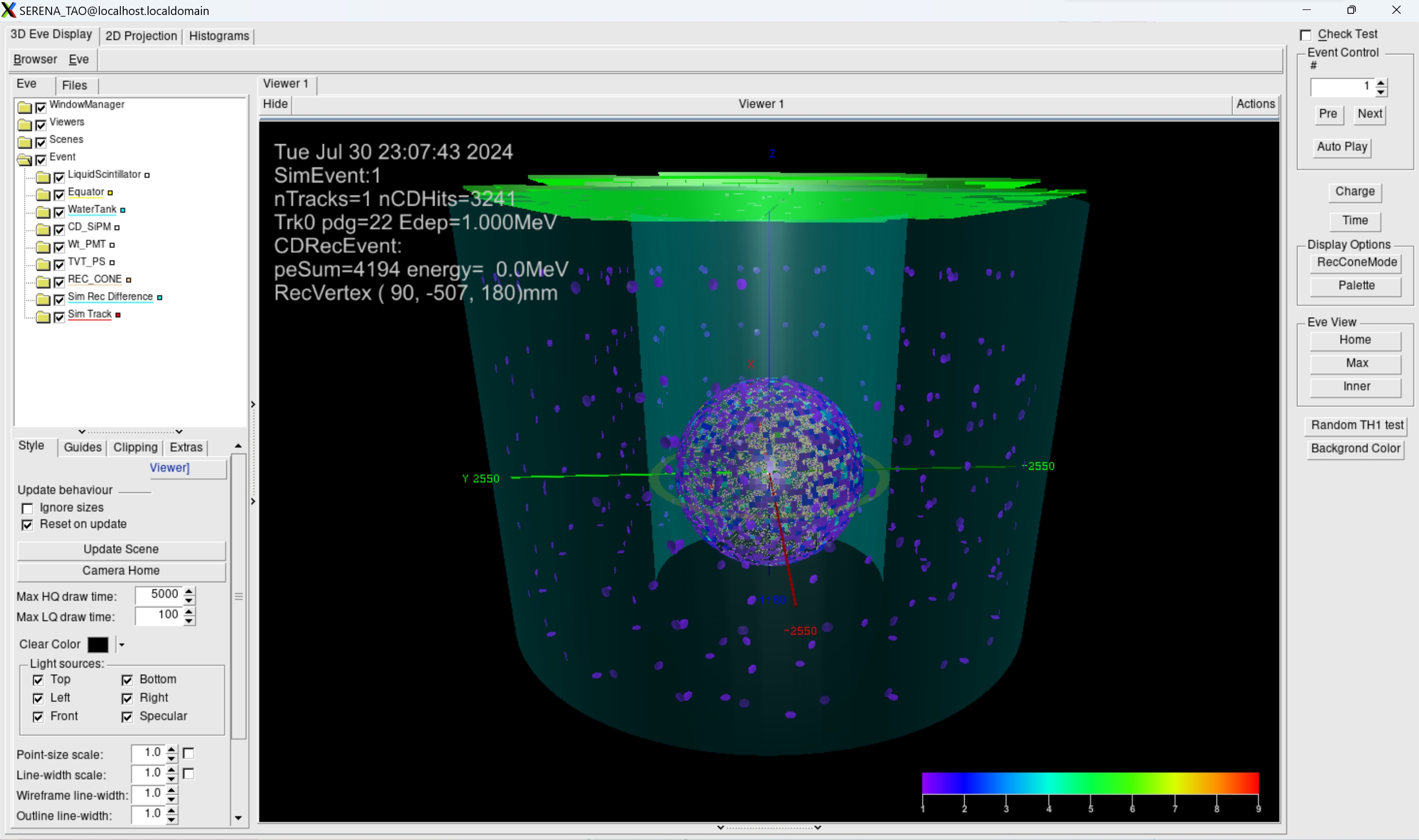}{\centering}
	\caption{GUI of the TAO event display software and display of a gamma event in the TAO detector.}
	\label{fig:GUI}
\end{figure*}

The Event Data Model~(EDM) offers a comprehensive array of event data information, which are represented as ROOT-based persistent data objects and flow between different stages and applications, such as SimEvent from detector simulation, ElecEvent from electronics, CalibEvent from calibration, and RecEvent from reconstruction output. The data is structured within a similar framework that aligns with the EDM in JUNO offline software~\cite{Li:2017zku, Lin:2022htc}.

The software structure of the event display shows that reading different data files through the event manager will distinguish which types of data these files contain. In the event display of TAO, various data files, such as SimEvent, CalibEvent, and RecEvent, are read simultaneously. The event data objects will be initialized with the corresponding ROOT EVE objects, such as the time and charge of fired PMT hits, reconstructed vertex and energy, Monte Carlo~(MC) truth information of each particle, and the photon propagation paths if provided. This information will be used in the subsequent development of the event display functions and display control from ROOT GUI. 

Since there are about 4,100 SiPMs, 300 PMTs, and 4 layers of TVT scintillators in TAO, constructing all the corresponding EVE objects is time-consuming, and it is impractical to perform such construction each time for each event. Instead, to display event by event more efficiently, the key is to construct all detector unit corresponding EVE objects only once at the initialization stage of event display software, and then change the visualization attributes of each detector unit object according to its fired status in a specific event. For example, the SiPMs can be displayed in different colors by the number of photons on this SiPM or by the time of the earliest arriving photon. The text information associated with this SiPM in EVE will also be updated from the EDM information. In this way, ROOT EVE can directly control the visualization of each event more efficiently, as shown in Fig.~\ref{fig:GUI}.

\subsection{EVE based visualization}

EVE is a ROOT module based on an experiment-independent part of the event display in the ALICE experiment\cite{Tadel:2008zz, ALICE:2008ngc}, which has been used to perform high-level event visualization, debugging of simulation and reconstruction code as well as for raw-data visualization.
The EVE package provides a standard platform for different experiments to visualize detector geometry and event information, combined with the GUI package to create an intuitive event display control window. 

As shown in Fig.~\ref{fig:GUI}, on the left side of the window is the default control panel of EVE, such as displaying detector and event information, and controlling the visual effects and settings. Some new features and functions specifically for TAO event display are developed and deployed with the corresponding widgets on the right panel, including event navigation, display of simulation and reconstruction, hit objects, and histograms. These functions are also linked to EVE visualization objects for interactive display.
%In the EVE window on the left side of the GUI, the buttons can control the display of detector geometry, mainly including CD, WT, and TVT. For event information, when the user clicks on the RecConeMode button, reconstruction, simulation, and their differences will also be displayed in the EVE window. For example, if you only want to see the position of reconstructed vertices, you can click on the Rec Cone button in the EVE window and disable other display buttons. In addition, based on the EVE package, we have implemented some other functions, such as displaying SiPM hits, including charge and time distribution. 
More details about visualization effects and functions will be discussed in Section~\ref{sec:vis}.

\subsection{GUI}

The GUI is an interface for interactive communication between the display and the users. The TAO event display GUI is implemented to display the detector geometry, event data, and the functional widgets for control. By clicking the functional buttons on the GUI panel, the users have access to control the visualization effects of the event and detector components. The GUI interface is constructed by combining the ROOT EVE and GUI packages. EVE provides the visualization functions, while GUI constructs the window panel and widgets to control these functions, as shown in Fig.~\ref{fig:GUI}.

The central window is the main body of the event display, including the display of the detector geometry structure, hits information, and photon tracks. It is constructed based on the OpenGL module for implementing data presentation and interaction layer. In the upper left corner, detailed event information is displayed to help the physicists understand the event in simulation and reconstruction, such as the identifier of a specific event, its production date and time, the total number of hits in this event, deposited energy, as well as reconstructed event vertex and the MC truth. 
%The color reference bar for each SiPM hit number and hit time is displayed in the lower right corner of the center section.

As shown in Fig.~\ref{fig:GUI}, both 3D display and 2D projection display are provided, which can be switched between different tabs in the upper left of the GUI window. In the left panel of the GUI window, some default functions are provided by ROOT EVE, such as the orientation of objects and setting the background color of the display window. In the right panel of the GUI window, some functional buttons are defined for event display, such as the event sequence control, charge and time display, event view mode, etc. The users can control the display of events in the detector through the GUI window and panels, as well as scan the events one by one to check more detailed event information for reconstruction algorithm tuning and physics analysis.

%%------------------------------------------
\section{Visualization}
\label{sec:vis}

This section is dedicated to presenting the realization of the visualization functionalities, including the visualization of detector geometry, hits, comparison of the reconstruction output, and 2D histograms.
%In sub-section A, the visualization of the TAO detector geometry is described in detail, including the central detector, water tank, and top veto track. The central detector is the primary focus of the research. We provide information on the distribution of hits on the central detector. This will be discussed in sub-section B. The sub-section C primarily describes the implementation and effect of the comparison of Rec-Cone mode for simulated and reconstructed information, including the import of files and the association of different data files. In sub-section D, we use the Aitoff method to project the SiPMs of the central detector onto a two-dimensional plane and provide the distribution of the 2D hits.

\subsection{Detector units}
%With the development of high-energy physics research, the structure of detectors is more complex. Visualizing the detector structure helps physicists more intuitively understand the overall layout of detectors and the distribution of tiny units.

As mentioned in Section~\ref{sec:structure}, all detector geometry information is converted from the GDML file and used as the input of the TAO event display. 
%It is a similar way to persistently store detector geometry in high-energy physics~\cite{You:2017zfr, Liang:2009zzb, You:2008}. The geometric information converted from the GDML file will be transferred and stored in a ROOT file and then used to construct the detector structure in the display, which provides a detailed description of the sub-detectors, such as SiPM and PMT positions, materials, shapes, and PS vertex positions.
However, it is challenging to visualize all the geometric details of every detector unit. 
%since the shape of the sub-detector unit and the peripheral water tank are both irregular. 
The graphical processing power requirements are usually beyond the computer hardware limits of most end-users.
On the other hand, the analyzers may not be interested in the exact shape of every detector unit like PMT but prefer to focus more on its spatial position and firing status.

Therefore, the optimization of detector units has been applied to the visualization. The units of detectors with complex structures are replaced by similar simple geometric shapes. For example, the SiPM and plastic scintillator are represented by the box objects~(TEveBox), and the PMTs are simplified into a set of cone objects, which makes detector visualization smoother on most end-user computers. With such implementation, the performance and interactive response speed of event display software has been greatly improved.

%--fig-- TAO detector
\begin{figure}[htbp]
    \centering
    \includegraphics[width=0.45\textwidth]{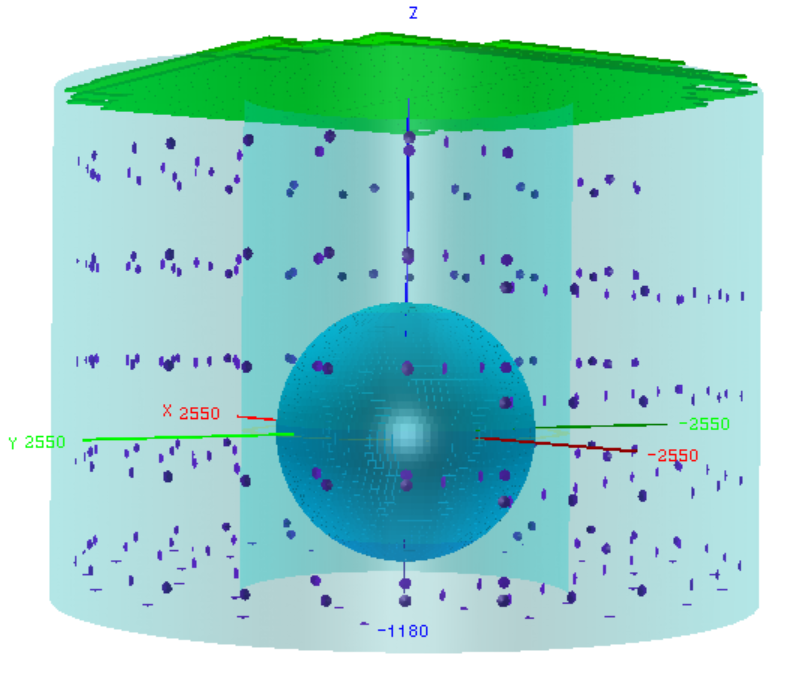}{\centering}
    \caption{Visualization of the TAO detector geometry in event display software. The light blue cylinder is the WT. The purple points in the WT are veto PMTs. The blue sphere is composed of SiPMs in the CD. The top green bars are four layers of plastic scintillators in the TVT. The numbers at the x, y, and z positions represent the set boundaries of the coordinate axes, with the unit being millimeters. This provides a visual understanding of the dimensions of the detector.}
    \label{fig:detector}
\end{figure}

As shown in Fig.~\ref{fig:detector}, the blue sphere is the visualization effect of CD, which is composed of 4,100 SiPMs. The light blue cylinder is the approximate shape of WT with 300 3-inch PMTs, which are represented as the little purple cones embedded in WT. The green TVT on the top consists of four layers of plastic scintillators that are approximately cuboid. 

Every detector object is selectable and pickable. For example, when the cursor moves over any SiPM or PMT, it is selected with a highlight visualization effect. Meanwhile, a text box pops up to show its spatial position and identifier, helping users get the detector unit information, as shown in Fig.~\ref{fig:charged}. The display properties of the detector units can be flexibly set by the EVE panel on the left. For example, the users can control whether to display a sub-detector or not by switching on/off the sub-detector on the left panel.

\subsection{Hits information}
\begin{figure}[htbp]
    \centering
    \includegraphics[width=0.45\textwidth]{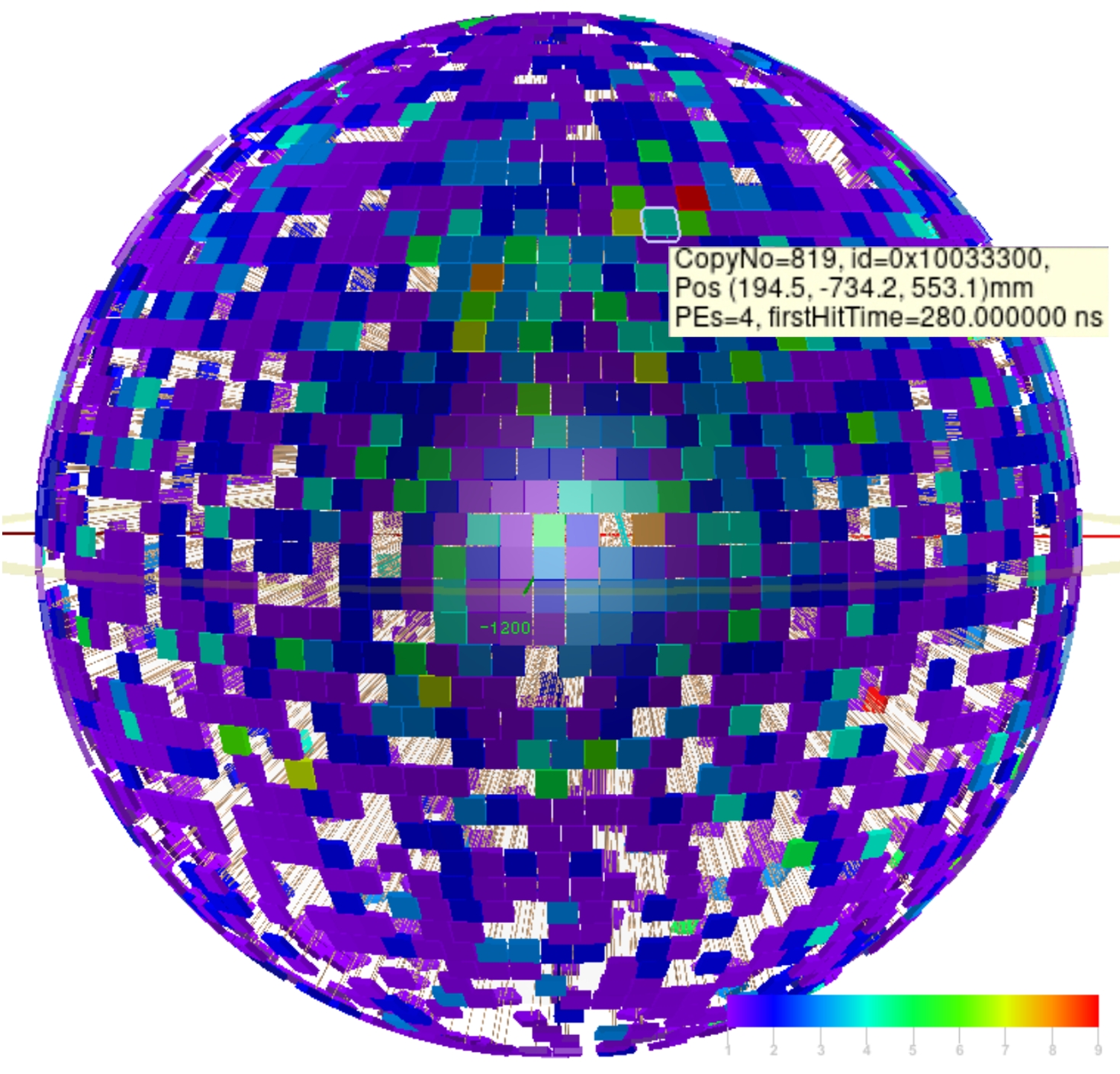}{\centering}
    \caption{The charge distribution of SiPM hits in an event, represented with different colors. The red SiPMs have more photon hits, while the blue and purple SiPMs have fewer hits, as illustrated in the legend.}
    \label{fig:charged}
\end{figure}

The critical function of visualization is to display the 
hits distribution. With about 4,100 sensitive SiPMs in CD, the photon detection efficiency is more than 50 percent. The event display software provides two modes to show the hit information of detector units: the number of photon hits on the SiPM~(charge) and the time of the first hit on the SiPM~(time). The charge distribution of an event in the CD is shown in Fig.~\ref{fig:charged}. The SiPMs are distinguished by their unique identifiers. The detector units that have been fired with hits will be visually represented with a specific color. Meanwhile, the detector units that have not been fired by any hit will be set invisible, allowing for clear visualization of the hit distribution in the whole detector.

%--charged distribution

The charge and time distribution play important roles in tuning the reconstruction algorithm. For example, the charge-weighted vertex reconstruction algorithm uses the hit charge distribution of SiPMs to reconstruct the event vertex. While the time distribution is used to calculate the spatial position of the maximum likelihood from an event vertex to every SiPM hit~\cite{JUNO:2020ijm, JUNO:2015sjr}. The charge and time distribution in the event display provides an intuitive way to help software developers understand the reconstruction algorithm. In addition, the charge and time distribution also help physicists understand the reaction processes while analyzing an event. 
%The difference in time distribution is slight due to the small radius of the central detector of TAO compared to the JUNO detector, while the difference in charge distribution is noticeable.

\subsection{Simulation and reconstruction information}
The MC simulation is critical for the design of detectors, helping physicists check the design performance of detectors and tune algorithms. 
%Monte Carlo simulations help physicists predict whether the sub-detector design meets the experiment's physical goals. 
The reconstruction algorithm uses the hits information received by the detector to reproduce the true information of an event as close to the real process as possible~\cite{Li:2021oos, Qian:2021vnh, Li:2022tvg}. 
When the experimental data is not available, MC simulation is a good approach to test the performance of the reconstruction algorithm.

An important function of the event display is to compare the differences between the reconstruction and MC truth in simulation to tune the accuracy of reconstruction algorithms and diagnose reconstruction results event by event to identify potential problems. For instance, in the context of Inverse Beta Decay~(IBD) events involving the positron's deposited energy vertex, discrepancies between the true vertex in MC and the reconstructed vertex can be effectively analyzed using the event display software. This enables the observation of distribution characteristics of such events, facilitating targeted improvement to the reconstruction algorithm. During event simulations, such as muon events in cosmic rays yielding a substantial number of photons, the event display software allows for the examination of charge and timing distributions within the SiPMs. The TAO event display software can simultaneously present various information about an event by reading different kinds of data, such as showing the reconstruction and simulation results together so that the users can visually compare the associations between them.

%--The cone of reconstruction
\begin{figure}[htbp]
    \centering
    \includegraphics[width=0.43\textwidth]{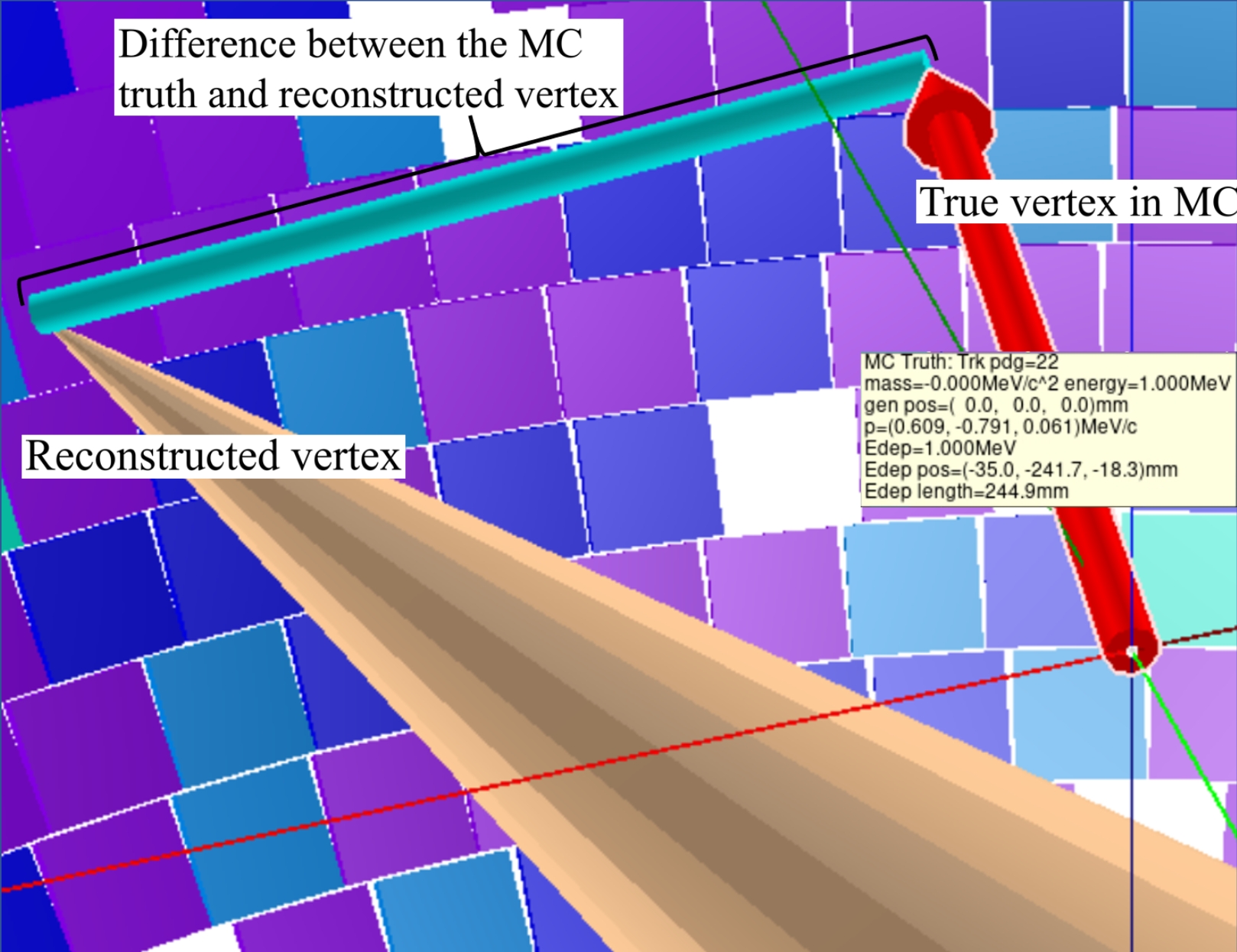}{\centering}
    \caption{Comparison of the reconstructed vertex with MC truth in a simulated positron event. 
    %The tail of the red arrow is the positron production vertex and the head of the arrow points to the position of positron annihilation from the MC truth.  The apex of the brown cone is the reconstructed event vertex. A blue line connecting the head of the red arrow and the apex of the cone illustrates the difference between reconstruction and MC truth. The text box exhibits the detailed MC truth information.
    }
    \label{fig:RecCone2}
\end{figure}
%Fig.~\ref{fig:RecCone2} shows the difference between the reconstructed vertex and the MC truth from a simulated positron event, which is extracted synchronously from the simulation and reconstruction output data files.

The difference between the reconstructed vertex and the MC truth from a simulated positron event, which is extracted synchronously from the simulation and reconstruction output data files, is shown in Fig.~\ref{fig:RecCone2}. The tail of the red arrow is the true production vertex position and the head of the red arrow points to the position of positron annihilation from the MC truth.  The apex of the brown cone is the reconstructed event vertex. A blue line connecting the head of the red arrow and the apex of the cone illustrates the difference between reconstruction and MC truth.

A shorter blue line indicates better reconstruction performance. For a perfect vertex reconstruction, the length of the blue line should be zero, and the reconstructed vertex merges to the same point as the true vertex.
The text box pops up to exhibit the detailed reconstruction or MC truth when the users move the cursor over the red or blue line to get more information.

\subsection{2D projection}

In addition to the 3D visualization of detector and event data, 2D projection is also helpful for data analysis. For example, the 2D event display of CMS~\cite{Kovalskyi:2010zz} and BESIII~\cite{BESIII:2009fln,Li:2024pox} can help physicists understand some data analysis processes more easily.

The 2D projection of CD SiPMs synchronously reads the geometry of the CD SiPM hits. ROOT EVE provides two default cross-section projection views, which are the $x$-$y$ view and $z$-$r$ view. However, it does not give much helpful information about the distribution of the hits since the CD is a symmetric sphere. The Aitoff projection method~\cite{AllenLafayette1998FlatteningTE} is implemented for the 2D projection view. It projects the whole surface of a sphere onto a 2D plane plot like spreading out the Earth into a flat surface.

%Fig.~\ref{fig:2D} shows the 2D charge distribution projection of the CD. 

\begin{figure}[htbp]
    \centering
    \includegraphics[width=0.43\textwidth]{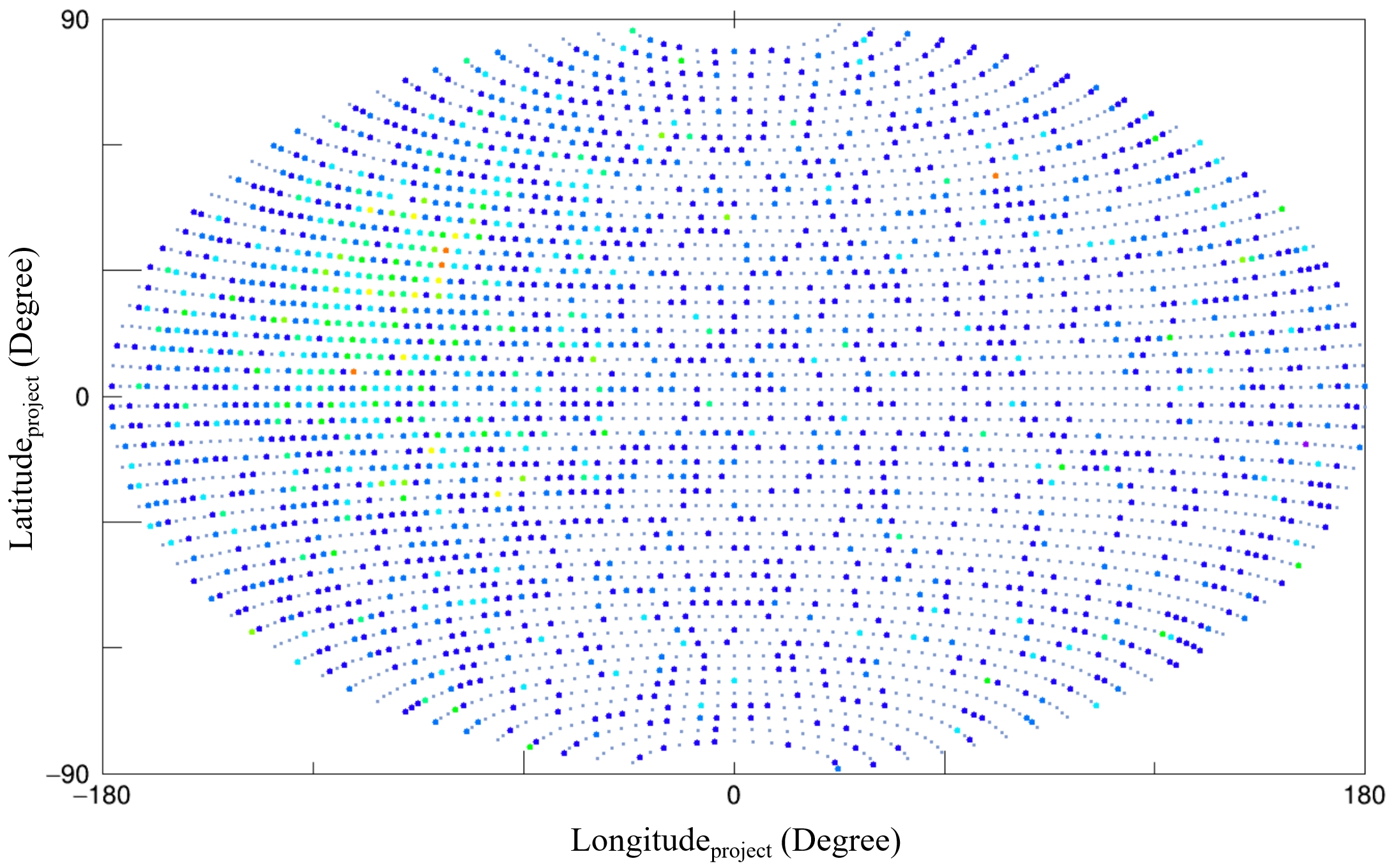}{\centering}
    \caption{The 2D projection view of an event in CD. Each maker represents a SiPM with different colors indicating the different number of photon hits on it. 
    %The color is closer to red representing more photons hit. The gray makers representing without photons hit the SiPM.
    }
    \label{fig:2D}
\end{figure}

The 2D charge distribution projection of the CD is shown in Fig.~\ref{fig:2D}. The solid markers with different colors represent the number of photons hitting the SiPM. The red color indicates more photons hitting the SiPM. The dots in grey represent the SiPMs without any hit on them. In a 2D projection view, the X and Y coordinates represent the longitude and the latitude of the CD sphere, respectively.

%--The cone of reconstruction

%%------------------------------------------
\section{Features and Applications}
\label{sec:fea}

The TAO event display software provides a concise and convenient interactive control interface between the users and the visualization of the event data.
Some features and applications of the visualization software are enumerated as follows.

\textit{Application in detector commissioning}. Event display plays an important role in detector commissioning by enabling the users to gain insights into the performance and operational status of the detectors. Through the analysis of events, the users can identify potential issues and evaluate detector performance, facilitating the debugging and repairing efforts. Furthermore, the visualization of the detector assists in optimizing detector performance, enhancing the efficiency and accuracy of data collection.

\textit{Visualization monitoring tool}. During the stage of detector running and experimental operation, the event display helps scientists to monitor and analyze experimental data in real-time~\cite{WOS:000462284000016, DayaBay:2017jkb, DayaBay:2019yxq}. It is utilized for the visualization and interpretation of particles interacting with the TAO detector, enabling researchers to swiftly discern critical features and patterns within the data. By displaying and analyzing the events from the reactor in real-time, the researchers can make timely modifications to the experimental parameters and data collection strategy, and optimize the detector performance to ensure high-quality experimental data taking.

\textit{Scan of rare signal events}. The TAO event display software supports two modes of visualization: 3D display and 2D projection, as shown in Fig.~\ref{fig:charged} and Fig.~\ref{fig:2D}. After the removal of a significant portion of the backgrounds which include the ambient neutron induced by cosmic ray muons and the radioactive backgrounds from the environments~\cite{Li:2022cjw, Li:2022wqc}, the rare signal events, such as reactor anti-neutrino events and other signal neutrinos~\cite{Berryman:2021xsi}, can be displayed within the detector in both visualization modes, facilitating an intuitive understanding of the physical processes to aid the analysis of rare events. In addition, the ultimate performance of the detector also relies on event reconstruction. To meet the requirements of the ultra-high energy resolution of TAO, it is necessary to fully utilize the charge and time information from SiPM and employ additional new technologies for reconstruction~\cite{Wei:2020yfs, Liu:2024cxo}. The event display provides support for the intuitive reconstruction of vertex and energy within the detector for comparison, as shown in Fig.~\ref{fig:RecCone2}. 
%It also facilitates the association of reconstructions with simulations, allowing for a comparison of the disparities between reconstructed vertices and simulation results, thereby demonstrating the quality of the reconstruction.

\textit{Nuclear power plant monitoring}. As a neutrino detector located very close to the Taishan NPP, TAO can also serve as an NPP monitoring apparatus. The visualization software will be instrumental in monitoring the operation not only for TAO but also for the NPP~\cite{Bowden:2008gu, NUCIFER:2015hdd, %WOS:000437273100001, WOS:000565778700001, WOS:000496964900002
}. Due to the advantage of monitoring a single reactor and superb energy resolution <2$\%$ at 1 MeV~\cite{JUNO:2020ijm}, the TAO experiment and its visualization tool provide a novel and unique approach for NPP monitoring in addition to the currently existing tools.

\textit{Future developments}. The event display based on ROOT is integrated into the offline software system as a module, commonly set up on servers running the LINUX operating system. Furthermore, JUNO has developed event display software using ROOT and Unity~\cite{You:2017zfr, Zhu:2018mzu}. In particular, Unity is adaptable for multi-platform use, including Windows, LINUX, MacOS, and web browsers. For TAO, the visualization will also be developed using Unity and Virtual Reality~(VR) applications~\cite{Yuan:2024sns}. More efforts are expected to implement the Unity-based event display for TAO, including the automatic detector geometry conversion~\cite{Huang:2022wuo}, as well as event data conversion from the ROOT format generated in the offline software to the format readable by Unity.

\section{Performance}
\label{Perfor}

In experimental evaluations conducted on a local computing system equipped with an Intel Core i7 CPU, the TAO event display software has demonstrated sufficient computational prowess to effectively process standard IBD events, achieving a display refresh rate exceeding 10 frames per second. The typical memory usage for visualizing IBD events in the TAO event display amounts to approximately 890~MB, encompassing the import and display of simulation, calibration, and reconstruction data files. Nevertheless, the operational efficiency of the software when accessed remotely is significantly contingent upon the latency experienced between the users and the computing server.

Scientific Linux is the predominant platform utilized for executing the TAO offline software, with the event display software having undergone rigorous testing and exhibiting robust performance on remote client systems such as Linux, Windows, and Mac. The event display software is incorporated into the TAO offline software framework, enabling streamlined interfacing with the online DAQ system for real-time monitoring and event visualization following requisite adaptations.

\section{Summary}
\label{sec:con}
The visualization technique serves as a valuable tool for HEP detector design, offline software development, and data taking. It can be utilized as an online data monitor tool during data acquisition and also plays an important role in enhancing the performance of detector simulation, reconstruction, and physics data analysis. 
As a detector situated in close proximity to the reactor, the visualization of JUNO-TAO can serve the research needs in neutrino physics and also aid users in monitoring the operation of NPP. 

With the construction of the JUNO-TAO experiment, the event display software has been successfully developed and more features are expected to be added in future updates. Furthermore, the Unity-based event display software is also under design and development, with significant potential for developing more advanced features and applications, such as online monitoring and VR programs. Following the successful application of visualization software for JUNO-TAO, visualization software based on ROOT and Unity will also be implemented in a broader range of future HEP experiments.

%--
\section*{Acknowledgements}
This work is supported by National Natural Science Foundation of China~(Grant No. 12175321, 11975021, 11675275),
the Strategic Priority Research Program of the Chinese Academy of Sciences under Grant No. XDA10010900. 
%the National College Students Science and Technology Innovation Project, 
%the Undergraduate Base Scientific Research Project of Sun Yat-sen University.

\printbibheading[heading=bibintoc]
\printbibliography[heading=none]
%\printbibliography

\end{document}